\UseRawInputEncoding
\documentclass[aps,prb,twocolumn,showpacs,psfig,superscriptaddress,longbibliography]{revtex4-2}
\usepackage{amsfonts}
\usepackage{mathrsfs}
\usepackage{amsmath}
\usepackage{color}
\usepackage{natbib}
\usepackage{textcomp}
\usepackage{graphicx}
\usepackage{bm}
\usepackage{amssymb}
\usepackage{braket}
\usepackage{xspace}
\usepackage{epstopdf}
\usepackage{dcolumn}
\usepackage{longtable}
\usepackage{multirow}
\usepackage[colorlinks=true, letterpaper=true, pdfstartview=FitV, linkcolor=blue, citecolor=blue, urlcolor=blue]{hyperref}
\usepackage{float}

\makeatletter

\newcommand{\Rmnum}[1]{\expandafter\@slowromancap\romannumeral #1@}
\makeatother

\begin{document}

\title{Family of New Binary Transition Metal Nitrides Superconductors}

 \author{Zheng-Wei Liao}
 \affiliation{School of Physical Sciences, University of Chinese Academy of Sciences, Beijing 100049, China}

 \author{Xin-Wei Yi}
 \affiliation{School of Physical Sciences, University of Chinese Academy of Sciences, Beijing 100049, China}

 \author{Jing-Yang You}
\email{phyjyy@nus.edu.sg}
\affiliation{Department of Physics, National University of Singapore, 2 Science Drive 3, Singapore 117551}

 \author{Bo Gu}
 \email{gubo@ucas.ac.cn}
 \affiliation{Kavli Institute for Theoretical Sciences, and CAS Center for Excellence in Topological Quantum Computation, University of Chinese Academy of Sciences, Beijng 100190, China}

 \author{Gang Su}
 \email{gsu@ucas.ac.cn}
 \affiliation{School of Physical Sciences, University of Chinese Academy of Sciences, Beijing 100049, China}
 \affiliation{Kavli Institute for Theoretical Sciences, and CAS Center for Excellence in Topological Quantum Computation, University of Chinese Academy of Sciences, Beijng 100190, China}

\begin{abstract}
Superconductivity in transition metal nitrides (TMNs) has been investigated for a long time, such as zirconium nitride (ZrN) with a superconducting transition temperature Tc of 10 K. Recently, a phase diagram has been revealed in ZrN$_x$ with different nitrogen concentrations, which is very similar to that of high-temperature copper oxide superconductors. Here, we study the TMNs with face-centered cubic lattice, where ZrN and HfN have been experimentally obtained, and predict eight new stable superconductors by the first-principle calculations. We find that CuN has a high Tc of 39 K with a very strong electron-phonon coupling (EPC) strength. In contrast to ZrN, CuN has softening acoustic phonons at the high symmetry point $L$, which accounts for its much stronger EPC. In addition, the highly symmetrical structure leads to topological protected nodal points and lines, such as the hourglass Weyl loop in $k_{x/y/z} = 0$ plane and Weyl points in $k_{x/y/z} = 2\pi / a$ plane, as well as quadratic band touch at $\Gamma$ point. CuN could be a topological superconductor. Our results expand the transition metal nitrides superconductor family and would be helpful to guide the search for high temperature topological superconductors.

\end{abstract}
\maketitle


Transition metal nitrides (TMNs), as a class of materials with excellent physical properties, such as high hardness and strength, strong corrosion resistance, high melting point, good chemical and thermal stability~\cite{Jhi1999,Zerr2003,Chhowalla2005,Crowhurst2006,Jin2020}, have attracted extensive attention and wide applications in various fields~\cite{Zou2016,Chatelain2020,Rivadulla2009,Wu2012,Ningthoujam2015}. In addition, many TMNs exhibit superconductivity~\cite{Papaconstantopoulos1985,You2021}. Since the 1930s, many TMN superconductors have been discovered. Nitrogen atoms in these superconductors provide strong bonding and large electron-phonon coupling (EPC), resulting in superconductivity~\cite{Matthias1952}. For instance, ZrN is a typical TMN superconductor, which displays the highest superconducting transition temperature Tc of 10.0 K among the IVB TMNs~\cite{Isaev2005,Weber1973}. As a hard metal nitride superconductor, ZrN is suitable for applications under extreme working conditions, while most materials with superior mechanical strength and hardness are semiconductors or insulators, which lack metallicity and superconductivity~\cite{Chen2005}. A recent theoretical study has shown that on the deformation path of high-symmetry crystallography [001], Tc of ZrN can reach 17.1 K under tensile strain~\cite{Lu2021}. Several other TMN superconductors have been reported including TiN with a maximal Tc of 6.0 K~\cite{Johansson1985}, HfN with a Tc of 8.8 K~\cite{Roberts1976}, VN with a Tc of 8-9 K~\cite{toth2014transition,Toth1966}, NbN with a Tc of 17-18 K~\cite{toth2014transition,Gurvitch1985,Pessall1968}, TaN with a Tc of 10.8 K~\cite{Reichelt1978}, and W$_2$N with a Tc of 1.3 K~\cite{toth2014transition}. MoN has been predicted to have the high Tc since 1981~\cite{Papaconstantopoulos1985,Pickett1981,Lengauer1988,Matthias1952,Zhao2000}. A high Tc of 30 K was theoretically predicted in MoN, but the low Tc of 5-14K were obtained in the experiments~\cite{Tsai1993,Cassinese2000,Nishi1987,Soignard2003,Papaconstantopoulos1984}, where the obtained MoN samples were not pure and often contaminated with Mo$_2$N, $\gamma$-Mo$_2$N, Mo or even MoO$_x$ phases.

In addition, studies have shown that the doping have a great impact on the superconductivity in TMNs. For example, the Tc of TiN$_{0.995}$, TiN$_{0.95}$, TiN$_{0.8}$, and TiN$_{0.55}$ are 6.0, 1.7, 1.5, and 1.2 K, respectively~\cite{Spengler1978}. Pure VN, 0.5\% B doped VN, 0.5\% La doped VN, 0.2\% B and La each co-doped VN, and 0.5\% B and La each co-doped VN have the Tc of 9.2, 8.0, 7.8, 8.2 and 5.6 K, respectively~\cite{Ningthoujam2002,Sudhakar2004,ningthoujam2004magnetic,Gajbhiye2006}. ZrN$_x$ with different nitrogen concentrations has been synthesized recently. Although ZrN$_x$ is generally believed to obey the Bardeen-Cooper-Schrieffer (BCS) superconducting mechanism, the phase diagram of ZrN$_x$ is very similar to that of high-Tc copper oxide superconductors, indicating the possible relation between high-temperature superconductivity and ZrN$_x$~\cite{chen2022emergence}.

In this article, we study the face-centered cubic (FCC) TMN in which ZrN and HfN have been experimentally reported. By substituting elements, eight new stable superconductors have been predicted through first-principle calculations. Interestingly, we found that CuN has a high Tc of 39 K with a very strong EPC of 3.099. In contrast to ZrN, CuN has softening acoustic phonons at the high symmetry point $L$, which accounts for its much stronger EPC. In addition, the highly symmetrical structure leads to topological protected nodal points and lines, such as the hourglass Weyl loop in $k_{x/y/z} = 0$ plane and Weyl points in $k_{x/y/z} = 2\pi / a$ plane, as well as quadratic band touch at $\Gamma$ point.  

\begin{figure}[!htbp]
	\centering
	\includegraphics[scale=0.55,angle=0]{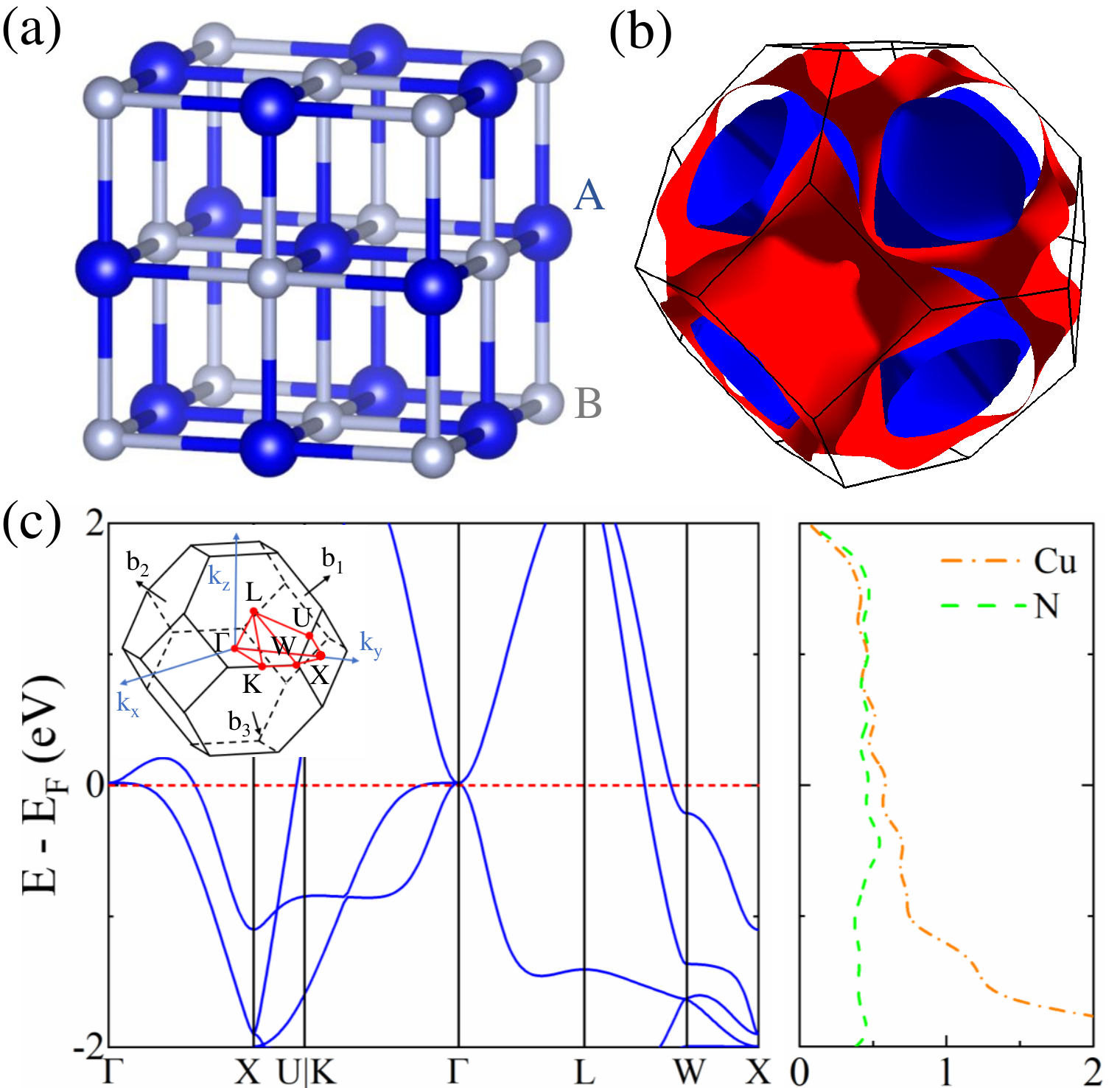}\\
	\caption{(a) The conventional cell of face-centered cubic lattice AB. (b) The Fermi surface, (c) electronic band structure and projected density of states (PDOS) of CuN.}\label{fig1}
\end{figure}
As shown in Fig.~\ref{fig1}(a), ZrN has a NaCl-type FCC lattice with the $Fm\overline{3}m$ space group. By substituting Zr with other transition metal elements and N with other VA group elements, we have obtained eight new structurally stable transition metal pnictides superconductors by the density functional theory (DFT) calculations, including CdN, CdP, CdAs, CuN, CuSb, HfP, HgSb, and ZnSb. Using the MicMillan-Allen-Dynes approach~\cite{Allen1975,McMillan1968} based on the BCS theory with a typical Coulomb repulsion value $\mu^*=0.1$, we calculate the Tc, EPC stength $\lambda$, and logarithmic average frequency $\omega_{log}$ of these superconductors as listed in Table~\ref{T-1}. In addition, we also calculate the Tc of two experimentally reported TMN superconductors ZrN and HfN, which are 10.93 and 9.1 K, respectively, close to the experimental Tc of 10 and 8.8 K. Among these stable FCC TMN superconductors, CuN has the strongest EPC, and the highest Tc of 39 K among other TMNs. In the following, we studied CuN as an example in details, and in the Supplemental Materials (SM) we provide the structural information, electronic bands, phonon spectra, density of states (DOS), and superconducting properties of other seven transition metal pnictides superconductors~\cite{SuplMat}.

The optimized lattice constant of CuN is 4.177 \AA. Fig.~\ref{fig1}(b) gives the Fermi surface of CuN, where red and blue surfaces represent the different bands crossing the Fermi surface, which obviously accords with the crystal symmetry of this structure. Fig.~\ref{fig1}(c) shows the band structure along the high-symmetry paths $\Gamma$-$X$-$U$|$K$-$\Gamma$-$L$-$W$-$X$ and projected DOS of CuN, which show the metallic property with the same contribution of Cu and N atoms near the Fermi level.

\begin{figure}[!htbp]
	\centering
	\includegraphics[scale=0.55,angle=0]{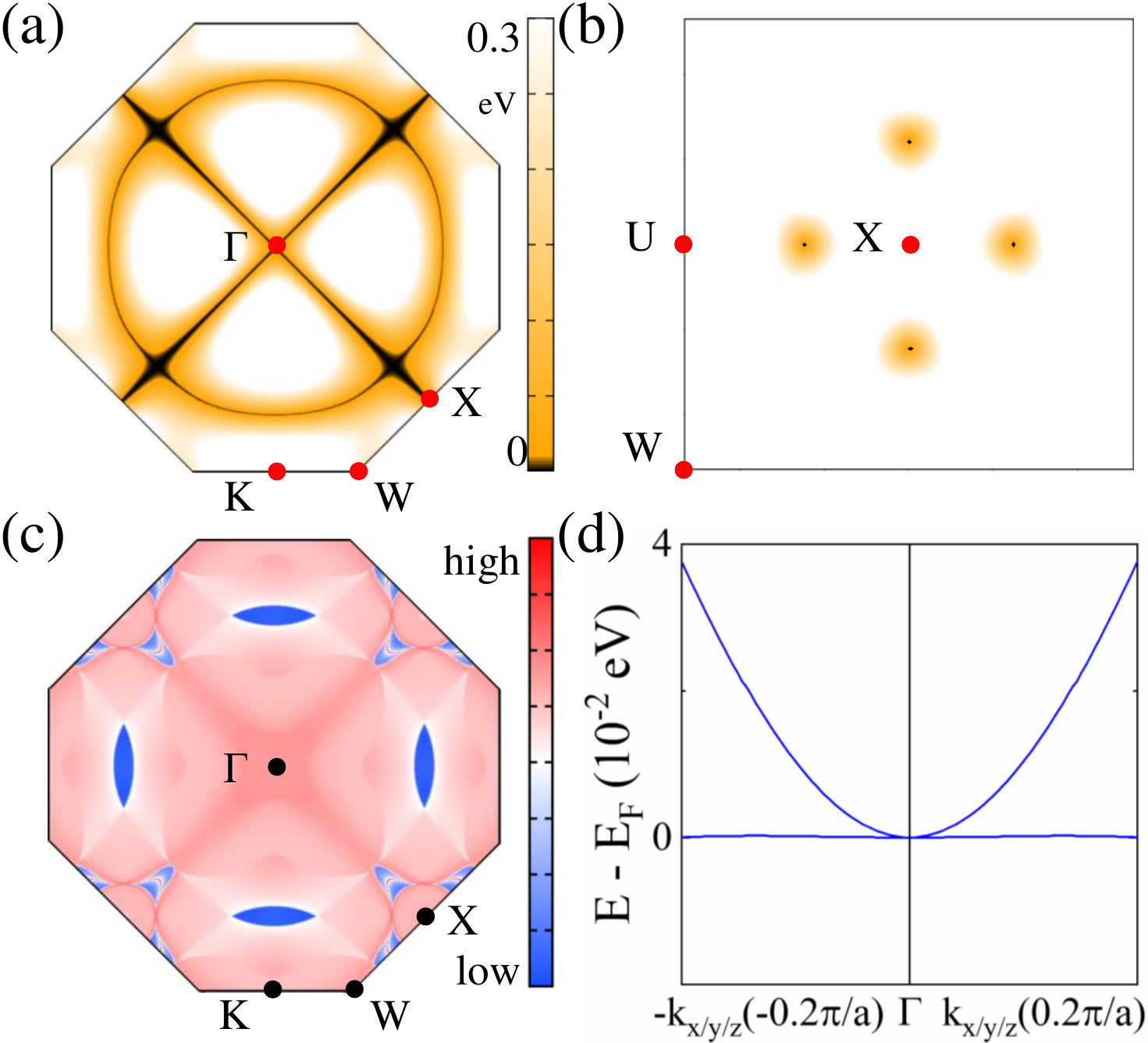}\\
	\caption{(a) Shape of the Weyl loop in $k_z = 0$ plane and (b) the distribution of Weyl points in $k_y = 2\pi / a$ plane for CuN. (c) The isoenergy slice at -0.8 eV projected on (001) surface of CuN. (d) The quadratic band dispersion around $\Gamma$ point.}\label{fig2}
\end{figure}

\begin{table}
	\renewcommand\arraystretch{1.25}
	\caption{The calculated total EPC $\lambda$, $\omega_{log}$, Tc of our proposed new stable compounds as well as two experimentally synthesized materials HfN and ZrN for comparision.}
	\begin{tabular}{l<{\centering}p{1.75cm}<{\centering}p{1.75cm}<{\centering}p{1.75cm}<{\centering}p{1.75cm}<{\centering}}
		\hline
		\hline
		& $\lambda$ & $\omega_{log}$ (K)  & Tc (K) & Tc$^{exp.}$ (K)  \\
		\hline 
		CdN          & 1.314     & 199.5               & 19.86  &  -       \\
		CdP          & 1.266     & 122.0               & 11.64  &  -       \\
		CdAs         & 1.426     & 94.9                & 10.29  &  -       \\
		CuN          & 3.099     & 124.2               & 39.29  &  -       \\
		CuSb         & 1.016     & 84.5                & 6.03   &  -       \\
		HfN          & 0.658     & 305.9               & 9.10   & 8.8~\cite{Roberts1976} \\
		HfP          & 0.670     & 151.8               & 4.74   &  -       \\
		HgSb         & 0.956     & 85.9                & 5.57   &  -       \\
		ZnSb         & 1.387     & 93.0                & 9.81   &  -       \\
		ZrN          & 0.657     & 375.3               & 10.65  & 10.0~\cite{Isaev2005,Weber1973}  \\
		\hline
		\hline
		
	\end{tabular}
	\label{T-1}
\end{table}	

\begin{figure}[!htbp]
	\centering
	\includegraphics[scale=0.48,angle=0]{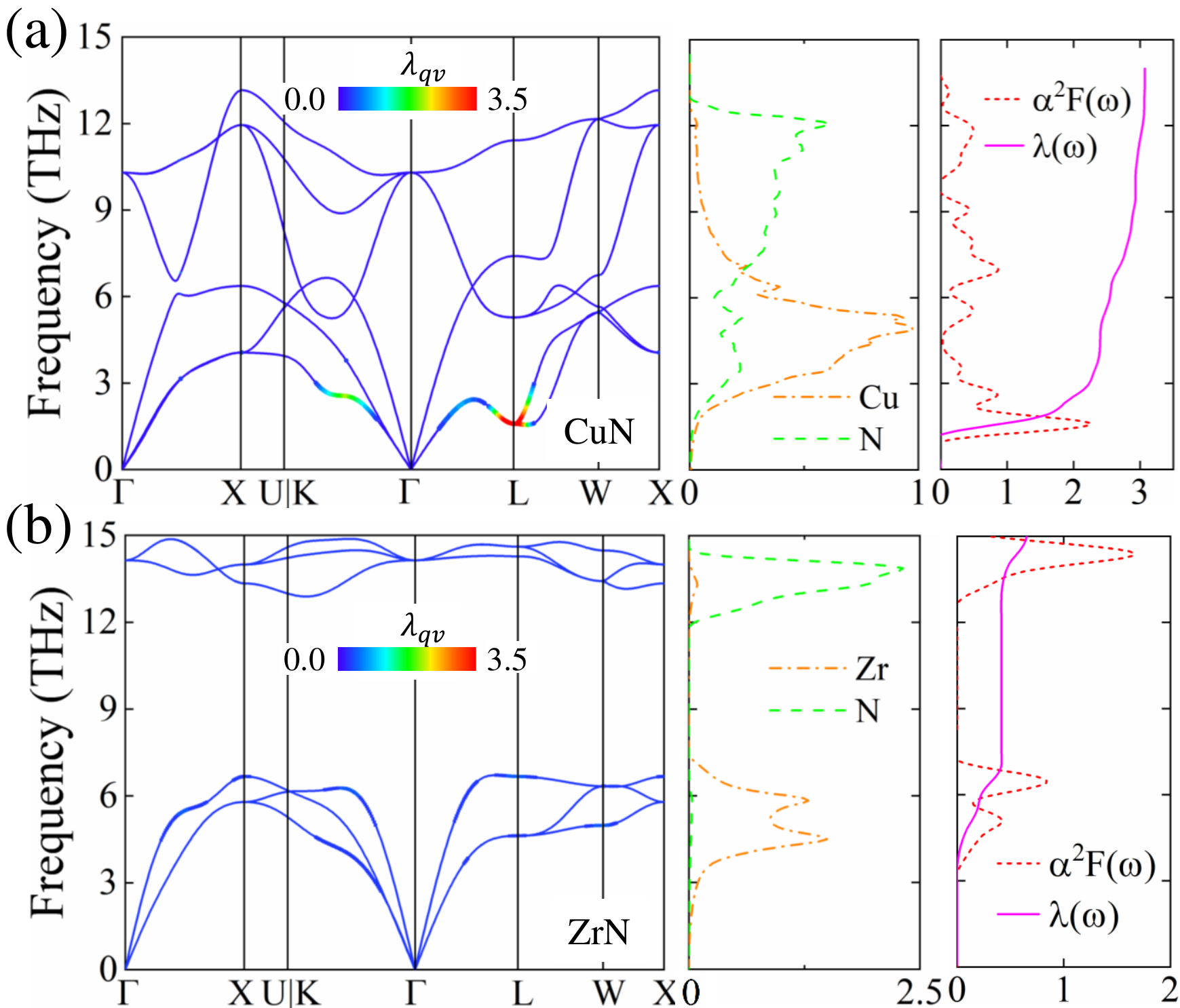}\\
	\caption{The phonon spectra weighted by the magnitude of EPC $\lambda_{q\nu}$, projected phonon density of states, Eliashberg spectral function $\alpha^2F(\omega)$, and cumulative frequency-dependent EPC $\lambda(\omega)$ of (a) CuN and (b) ZrN. }\label{fig3}
\end{figure}

Because of the highly symmetric crystal structure, this material can exhibit many interesting topological properties in the band structure~\cite{You2021a}. 
Fig.~\ref{fig2}(a) shows the shape of the nodal loop obtained from the DFT calculations in the $k_z = 0$ plane within the Brillouin zone (BZ). Similarly, other two nodal loops appearing on $k_x = 0$ and $k_y = 0$ planes can also be obtained by symmetry. In Fig.~\ref{fig2}(c), we plot the constant energy slice at -0.8 eV, which cuts through the drumhead states, forming a few arcs.
In Fig.~\ref{fig1}(c), we can find another band crossing point along $U$-$X$ path at about -1 eV. It is a nontrivial band crossing, leading to a nodal Weyl point. Since the fourfold rotation symmetry along $\Gamma$-$X$, there should exist another three Weyl points in the $k_y = 2\pi / a$ plane, which has been shown in Fig.~\ref{fig2}(b). In addition, the band dispersion around $\Gamma$ points is quadratic along all three directions in $k$ space as shown in Fig.~\ref{fig2}(d), indicating that CuN possesses the quadratic contact point $\Gamma$ protected by the crystalline symmetry~\cite{You2020}.

In Fig.~\ref{fig3}, we study the phonon spectra weighted by the magnitude of EPC $\lambda_{q\nu}$, projected phonon DOS, Eliashberg spectral function $\alpha^2 F(\omega)$, and cumulative frequency-dependent EPC $\lambda(\omega)$ for CuN and ZrN. For CuN, the vibration modes can be divided into two parts: low-frequency modes (< 8 THz) contributed by both Cu and N atoms and high-frequency modes (> 8 THz) dominated by vibration of N atoms. For ZrN, the vibration modes can also be divided into two parts: low-frequency modes and high-frequency modes dominated by Zr and N atoms, respectively. Between the two frequency ranges of ZrN, there is a 4.5 THz phonon gap. It is worth noting that at the high symmetry point L, the acoustic phonons of CuN exhibit obvious softening, accounting for about 83\% of the total EPC ($\lambda=3.099$) below 6 THz, while ZrN has no softening acoustic phonons, and the vibration modes of Zr atom below 7.5 THz only account for 63\% of the total EPC ($\lambda=0.657$) below 7.5 THz. Softening acoustic phonons are the main source for the large EPC in CuN.

\begin{table}[htbp]
	\renewcommand\arraystretch{1.25}
	\caption{The DOS at Fermi level $N(E_F)$ (in unit of states/spin/eV/cell), volume of primitive cell $V$ (\AA$^3$), $\omega_{log}$ (K), $\lambda$, and Tc (K) for CuN and ZrN as a function of pressure.}
	\label{T-2}
	\begin{tabular}{l<{\centering}p{1.2cm}<{\centering}p{1.2cm}<{\centering}p{1.2cm}<{\centering}p{1.2cm}<{\centering}p{1.2cm}<{\centering}p{1.2cm}<{\centering}}
		\hline 
		\hline
		&$P$(GPa)   &$N(E_F$) &$V$(\AA$^3$)  &$\omega_{log}$(K)   &$\lambda$ &Tc(K) \\
		\hline 
		\multirow{5}{*}{CuN}
		&0    &8.19  &18.22   &124.2     &3.099     &39.31          \\
		&1    &8.15  &18.14   &154.3     &2.377     &37.23        \\
		&2    &8.10  &18.05   &180.1     &1.989     &35.83       \\
		&3    &8.07  &17.97   &202.0     &1.737     &34.45      \\
		&4    &8.03  &17.89   &217.7     &1.587     &33.38          \\
		&5    &7.99  &17.81   &229.1     &1.488     &32.45             \\
		\hline
		\multirow{5}{*}{ZrN}
		&0    &4.47  &24.24   &375.3     &0.653     &10.93          \\
		&1    &4.45  &24.15   &377.3     &0.649     &10.80        \\
		&2    &4.43  &24.05   &381.5     &0.643     &10.66       \\
		&3    &4.41  &23.96   &384.7     &0.639     &10.50       \\
		&4    &4.39  &23.87   &387.9     &0.633     &10.39          \\
		&5    &4.38  &23.78   &391.0     &0.629     &10.26             \\
		\hline \hline
	\end{tabular}
\end{table}

\begin{figure}[!htbp]
	\centering
	\includegraphics[scale=0.62,angle=0]{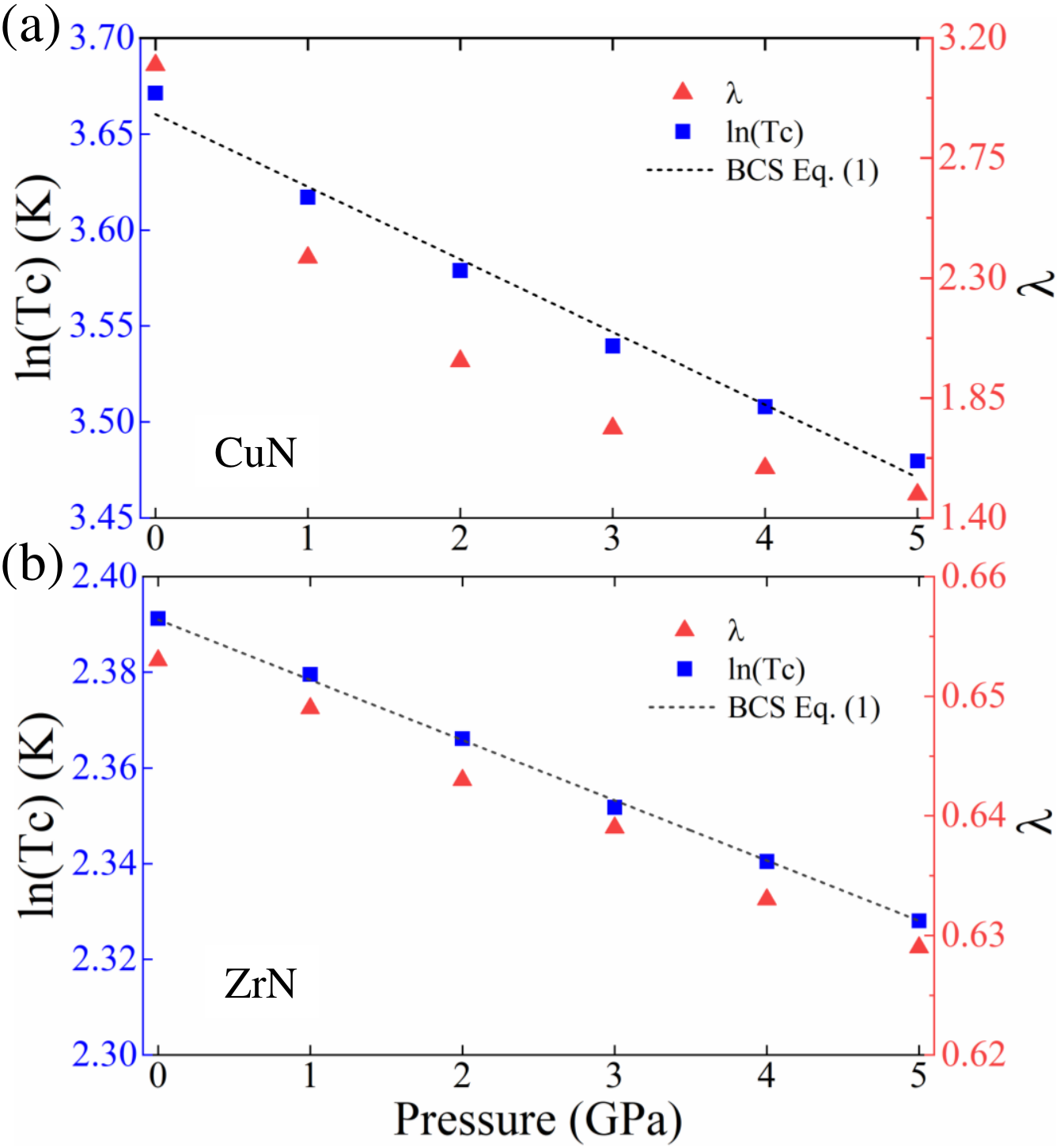}\\
	\caption{Pressure dependent logarithmic Tc and $\lambda$ of (a) CuN and (b) ZrN. The dash line is the fitting according to BCS theory.}\label{fig4}
\end{figure}

Fig.~\ref{fig4} gives pressure dependence of logarithmic Tc and EPC $\lambda$ of CuN and ZrN, respectively. Their detailed calculation results are listed in Table~\ref{T-2}. It can be seen that with increasing pressure, the EPC $\lambda$ of both CuN and ZrN decrease, resulting in the decrease of Tc. The calculated $d$Tc/$dP$ of ZrN is -0.133 K/GPa, which is close to the experimental result of -0.17 K/GPa~\cite{Chen2004}.
Based on the BCS theory, Tc($P$) obeys the following relationship\cite{schilling2007high}:
\begin{equation}\label{eq-Tc}
	\frac{d \rm ln Tc}{d {\rm ln} V}=-B\frac{d \rm ln Tc}{d P}=-\gamma+\Delta \{\frac{d {\rm ln} \eta}{d {\rm ln} V}+2\gamma\},
\end{equation}
where $B$ is the bulk modulus, $\gamma$ $\equiv$ -$d$ln$\langle \omega \rangle$/$d$ln$V$ is the Gr\"{u}neisen parameter, $\eta$ $\equiv$ $N(E_F)\langle I^2\rangle$ is the Hopfield parameter~\cite{Hopfield1971} with $\langle I^2\rangle$ the square of the electron-phonon matrix element averaged over the Fermi surface, and $\Delta$ $\equiv$ 1.04$\lambda$(1+0.38$\mu^*$)[$\lambda$-$\mu^*$(1+0.62$\lambda$)]$^{-2}$. Because the first term on the right hand side of Eq.~(\ref{eq-Tc}) is smaller than the second term~\cite{Chen2002}, the sign of $d$Tc/$dP$ is mainly determined by the relative magnitude of the two terms in curly bracket in Eq.~(\ref{eq-Tc}). The Gr\"{u}neisen parameter ($\gamma$) can  be directly obtained from Table~\ref{T-2}, and the Hopfield term can be determined by Eq.~(\ref{eq-Tc}). It can be observed that our calculated results for ZrN are in perfect agreement with Eq.~(\ref{eq-Tc}). For CuN, our calculated $d$lnTc/$dP = -0.0304$ K/GPa from Eq.~(\ref{eq-Tc}) is also comparable with the fitting slope $-0.035$ K/GPa in Fig.~\ref{fig4}(a). 

In summary, we have found eight new stable transition metal pnictides superconductors by the first-principle calculations. Among these superconductors, CuN is found to have the strongest EPC of 3.099 and the highest Tc of 39 K. Considering the high symmery crystal structure, we have obtained the quadratic contact point at $\Gamma$, where the band dispersion is quadratic along all three directions in $k$ space. Besides, Weyl loop in $k_{x/y/z} = 0$ plane and Weyl points in $k_{x/y/z} = 2\pi /a$ plane have also been uncovered, indicating the topological properties of CuN. In contrast to ZrN, we find that the softening acoustic phonon models in CuN are responsible for the much higher EPC strength and the much higher Tc. By studying the relationship between Tc and pressure, we find that both ZrN and CuN match well with the BCS superconducting mechanism. Our results not only predict CuN having the highest Tc in TMN superconductors, but also offer new materials with both superconductivity and novel topology in band structures that would be helpful for studying majorana zero modes in topological quantum computation.

This work is supported in part by the National Key R\&D Program of China (Grant No. 2018YFA0305800), the Strategic Priority Research Program of the Chinese Academy of Sciences (Grant No. XDB28000000), the National Natural Science Foundation of China (Grant No.11834014), and Beijing Municipal Science and Technology Commission (Grant No. Z118100004218001). B.G. is supported in part by the National Natural Science Foundation of China (Grants No. 12074378 and No. Y81Z01A1A9 ), the Chinese Academy of Sciences (Grants No. YSBR-030, No. Y929013EA2), the Strategic Priority Research Program of Chinese Academy of Sciences (Grant No. XDB33000000), and the Beijing Natural Science Foundation（(Grant No. Z190011).


%

\end{document}